\documentclass[twocolumn]{revtex4}
\usepackage[]{hyperref}

\usepackage[utf8]{inputenc}
\usepackage[T1]{fontenc}
\usepackage{fixltx2e}
\usepackage{graphicx}
\usepackage{longtable}
\usepackage{float}
\usepackage{wrapfig}
\usepackage{rotating}
\usepackage[normalem]{ulem}
\usepackage{amsmath}
\usepackage{textcomp}
\usepackage{marvosym}
\usepackage{wasysym}
\usepackage{amssymb}
\usepackage{hyperref}
\usepackage{xcolor}
\usepackage{graphicx}
\usepackage{tikz}
\usepackage{fp}
\usepackage{relsize}
\usepackage{fancybox}
\usetikzlibrary{decorations.pathmorphing, patterns,angles,quotes}
\usepackage{siunitx}
\usepackage{amsmath}

\newcommand{\sub}[1]{_{#1}^{\phantom{\dagger}}}
\newcommand{\super}[1]{^{#1}_{\phantom{\dagger}}}
\newcommand{\ad}[1]{a_{#1}^{\dagger}}
\newcommand{\an}[1]{a_{#1}^{\phantom{\dagger}}}

\newcommand{\bd}[1]{b_{#1}^{\dagger}}
\newcommand{\bn}[1]{b_{#1}^{\phantom{\dagger}}}

\newcommand{\cd}[1]{c_{#1}^{\dagger}}
\newcommand{\cn}[1]{c_{#1}^{\phantom{\dagger}}}

\newcommand{\bad}[1]{\bar a_{#1}^{\dagger}}
\newcommand{\ban}[1]{\bar a_{#1}^{\phantom{\dagger}}}
\newcommand{\fd}[1]{f_{#1}^{\dagger}}
\newcommand{\fn}[1]{f_{#1}^{\phantom{\dagger}}}

\newcommand{\hc}{\mbox{H.~c.}}

\newcommand{\ncr}{\nonumber\\}

\newcommand{\rhs}{right-hand side}

\newcommand{\guniv}{G_{S}^{\phantom\dagger}}
\newcommand{\gtwo}{\mathcal{G}_2^{\phantom\dagger}}
\newcommand{\ie}{i.e.,}

\newcommand{\etal}{{\em et al.\/}}

\newcommand{\grp}{Ref.~\onlinecite{GRP+08:246601}} 
\newcommand{\vw}{W}

\usepackage{amsmath}

\begin{document}

\title{How sharply does the Anderson model depict a single-electron transistor?}

\author{Krissia Zawadzki}
\author{Luiz N. Oliveira}
\affiliation{Departamento de F\'isica e Ci\^encia Interdisciplinar \\
Instituto de F\'{\i}sica de S\~ao Carlos, University of S\~ao Paulo\\
  Cx.~Postal~369, 13560-970 S\~ao Carlos, SP, Brazil}

\begin{abstract}
The single-impurity Anderson model has been the focus of
  theoretical studies of molecular junctions and the single-electron
  transistor, a nanostructured device comprising a quantum dot that
  bridges two otherwise decoupled metallic leads. The low-temperature
  transport properties of the model are controlled by the ground-state
  occupation of the quantum dot, a circumstance that recent
  density-functional approaches have explored. Here we show that the
  ground-state dot occupation also parametrizes a linear mapping
  between the thermal dependence of the zero-bias conductance and a
  universal function of the temperature scaled by the Kondo
  temperature. Careful measurements by Grobis and co-workers are very
  accurately fitted by the universal mapping. Nonetheless, the dot
  occupation and an asymmetry parameter extracted from the same mapping are
  relatively distant from the expected values. We conclude that
  mathematical results derived from the model Hamiltonian reproduce
  accurately the universal physical properties of the device. In
  contrast, non-universal features cannot be reproduced
  quantitatively. To circumvent this limitation, \emph{ab initio}
  studies of the device at high energies seem necessary, to accurately
  define the model Hamiltonian. Our conclusion reinforces findings by
  Gross and coworkers, who applied time-dependent density-functional
  theory to show that, to describe the low-energy properties of
  molecular junctions, one must be able to describe the high-energy regime.  
\end{abstract}

\maketitle

\section{Introduction}
\label{sec:1}
Molecular junctions and analogous elementary nanostructured devices have motivated a great deal of
experimental and theoretical research \cite{2005BGG+,2009NaB,2013StL,2018ThF030901}. Archetypical
among such systems is the single-electron transistor (SET), a quantum dot or molecule (\emph{dot},
for briefness) bridging two otherwise decoupled 2D electron gases or metallic leads (\emph{leads})
\cite{GR87:452,NgL88:1768,GSM+98.156,GGK+98.5225}. That the single-impurity Anderson Hamiltonian
would model the transport properties of the device was realized well before the first SET was
manufactured. Two corollaries emerged. First, in view of the universal properties of the Anderson
Hamiltonian, quantitative interpretation of experimental data was envisaged.  Second, given that the
dot occupation controls the ground-state transport properties, the model invited Density-Functional
Theory (DFT) treatment.  With the invitation, alas, came a challenge.

A formidable barrier faces density-functional theorists interested in molecular junctions or SETs. A
crossover separates the high-energy properties from the low-energy properties. The crossover is
refractory to perturbative treatment. Only special methods can treat it.

At first, difficulties other than the crossover attracted attention
\cite{2010MeN216408}. DFT is centrally concerned with the ground
state; research was therefore focused on the low-temperature behavior,
the transport properties being computed via Landauer-B\"uttiker
formalism \cite{1986But1761}.  With a view to developing trustworthy
approximations for the exchange-correlation functional, accurate
special results such as Density-Matrix Renormalization-Group data
\cite{2008ScE086401}, the Friedel sum rule \cite{PhysRevB.85.115409},
the behavior of an isolated impurity in the low-temperature limit
\cite{2011StK216401}, and Bethe-Ansatz results for the ground-state
occupancy of the Anderson-model impurity
\cite{BLB2012_066801,2012LBBp155117} were invoked.

In this charged environment, the obstacle that lay ahead might have been disregarded, had
Hardy and collaborators not issued the heads up. In an inspiring report, they showed that,
unlike static DFT, the time-dependent formalism (TD-DFT) can climb the crossover
\cite{2010KSKp236801}, \ie\ it can describe the high-energy region, from which the system
inherits its low-energy properties. In the wake of this news came a sequence of
developments that opened inroads towards the solution of the non-equilibrium problem
\cite{2011StK216401,2016KuS241103,2017KuS,2018JaK00255}.

This remarkable progress notwithstanding, some of the work that was done after
Ref.~\cite{2010KSKp236801} came to light indicates that Hardy's message has not come
across clearly.  Another shot seems in order.

The complexity of the crossover and its relation to high- and to the low-temperature
properties can be perceived from another perpective, rooted in physical
considerations. The crossover is due to the formation of the Kondo cloud. At high
energies, if the gate voltage attracts an odd number of electrons, the dot acquires a
magnetic moment. The dot moment is antiferromagnetically coupled to the moments of the
nearby lead electrons. As the temperature $T$ is lowered past a characteristic temperature
$T\sub{K}$ (the Kondo temperature, typically of the order of \SI{1}{K}), a cloud arises in the
leads that couples with the dot spin to form a singlet. Below $T\sub{K}$, the entanglement
between the dot and lead electrons allows ballistic conduction across the device.

The cloud is large. The correlation length grows as the temperature is reduced, \ie\ as
the Hamiltonian crosses over from the high- to the low-temperature regimes, and may exceed
$\SI{1}{\mu}$. Such long lenghts introduce non-local effects that simple approximations to
the static exchange-correlation functional are unlikely to capture. Discussion of the DFT
approach from a strategical perspective seems therefore warranted. Along that line of
reasoning, the crossover merits special attention.

We find convenient mathematical expression of the transition in the thermal dependence of
the transport properties of the model Hamiltonian. Here, we show that, even when external
potentials are applied to the leads, temperature-dependent zero-bias electrical
conductance maps linearly onto a universal function. Like the low-energy properties, the
mapping is parametrized by the ground-state expectation value for the dot occupation.

How reliable is this universal relation? The mapping has been thoroughly checked against Numerical
Renormalization-Group data \cite{YSO2009:235317}. The accurate experimental data reported by Grobis
\etal\ \cite{GRP+08:246601} pose a more trying test. Reference~\cite{GRP+08:246601} systematically
applied a sequence of gate voltages at various temperatures to span the Kondo regime.  We fit the
thermal dependence at each gate voltage with the universal expression and extrapolate the
experimental data to temperatures $T\gg T\sub{K}$ and the $T\ll T\sub{K}$. The extrapolations
determine the conductance at temperatures well outside the experimentally accessible thermal
range. They also determine the Kondo temperature and ground-state dot occupations as functions of
the gate potential.  The comparison shows that the universal properties of the model Hamiltonian
reproduce the experimental data quantitatively. By contrast, the nonuniversal properties agree but
qualitatively with parameters derived from the experimental results.

We conclude that accurate diagonalization of a simplified Hamiltonian is sufficient to describe the
crossover to the low-temperature regime, while \emph{ab initio} treatment is necessary to describe
other aspects of the experiment. To draw attention to the practical implications of this conclusion,
we recall the renormalization-group argument showing that the high-energy
spectrum is approximately reproduced by a single-particle Hamiltonian devoid of characteristic
energies \cite{Wi75:773}. The similar spectra open an opportunity for DFT descriptions of the device
at high energies, which can be combined with a nonperturbative treatment of the crossover to reach
the low-temperature regime.

The paper is structured as follows. Section \ref{sec:50} defines the model Hamiltonian and cursorily derives the general expression for the zero-bias conductance. Section \ref{sec:3} discusses the characteristics energies of the model and the special regimes they define and then discusses the mapping of the conductance to a universal function. Section \ref{sec:24} shows that the mapping fits experimental data quantitatively and compares the resulting dot occupancies with expected values. A summary section caps the text. 

\section{Conductance}\label{sec:50} 
Our analysis slightily extends the work of Yoshida \etal\ \cite{YSO2009:235317}, who showed that the thermal
dependence of the SET conductance in the Kondo regime maps linearly to a universal function of the
temperature scaled by the Kondo temperature. The linear coefficient is a trigonometric function of
the ground-state phase-shifts induced by the screening of the dot moment. Here, we will allow for an
external potential applied to the leads and take
advantage of Friedel's sum rule to relate the phase shifts to the $T\ll T\sub{K}$ and
$T\gg T\sub{K}$ occupations of the dot orbital.

\begin{figure}[!ht]
  \centering
  \includegraphics[width=0.5\textwidth]{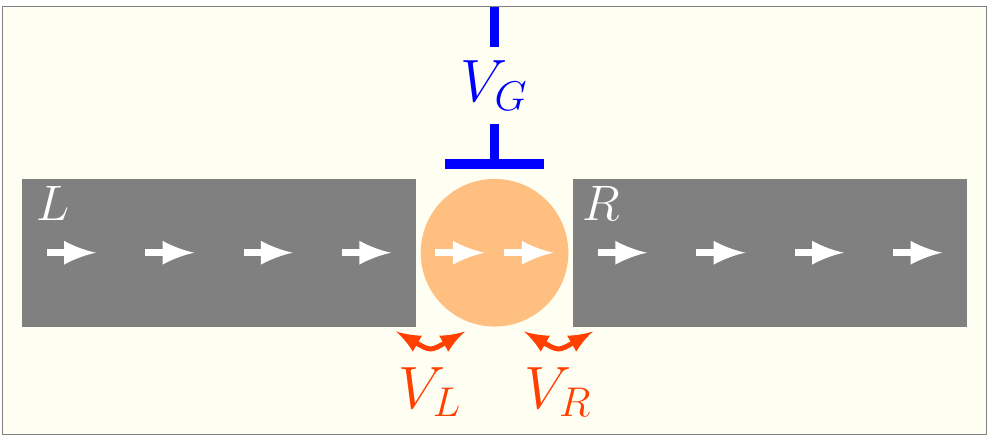}
  \caption{(Color online) Single-electron transistor. The quantum dot is asymmetrically coupled to
    the left ($L$) and the right ($R$) leads, with couplings $V_{L}$ and $V_{R}$,
    respectively. The gate potential $V_G$ controls the dot energy. The arrows indicate the
    direction of conduction}
  \label{fig:1}
\end{figure}

The schematic drawing in Fig.~\ref{fig:1} defines the model. (For a micrograph of the modeled
device, see Figure~1(a) in Ref.~\cite{GRP+08:246601}.)  The quantum dot, at the center of the
figure, is coupled to the left ($L$) and the right ($R$) leads with couplings $V_{L}$ and $V_{R}$,
respectively.  The dot-level energy, and hence the dot occupancy are controlled by the gate
potential $V_{G}$.

The Anderson Hamiltonian modeling the device in Fig.~\ref{fig:1} reads
\begin{align}
  \label{eq:1}
  H\sub{A}&= H_d + \sum\sub{k,\alpha=L,R}\epsilon\sub{k}\cd{k\alpha}\cn{k\alpha}
+\dfrac{\vw}{N}\sum\sub{k,q,\alpha=L,R}\cd{k\alpha}\cn{q\alpha}\ncr
  &+\sum\sub{\alpha=L,R}\dfrac{V\sub{\alpha}}{\sqrt{N}}(\cd{d}\cn{k\alpha}+\hc).
\end{align}
with implicit spin sums throughout. As usual, the dot Hamiltonian $H_d$ comprises a Coulomb
repulsion $U$ and a dot energy $V\sub{G}$, defined by the gate potential.  The two structureless
conduction bands in the first term on the \rhs\ represent the left- ($\alpha=L$) and the right-hand
$(\alpha=R)$ leads. The normalized sum $(1/\sqrt{N})\sum\sub{k\alpha}\cn{k\alpha}$ defines the Wannier
state in lead $\alpha$ to which the dot level $\cn{d}$ is coupled.

The second term on the \rhs\ of Eq.~(\ref{eq:1}) represents equal external potentials applied to the
same Wannier states. The potentials must be identical to maintain equilibrium, but the application
to the orbitals coupled to the dot is by no means restrictive. Renormalization-group theory proves
that substitution of a momentum-dependent form
$\sum\sub{k,q,\alpha}W\sub{kq}\cd{k\alpha}\cn{q\alpha}$ for the second term on the \rhs\ would only
add irrelevant terms to the Hamiltonian \cite{KWW80:1003,Wi75:773}.

Such irrelevant operators would contribute to physical properties at high energies. For decreasing
temperatures, however, the contribution would decay in proportion to $T$, or more rapidly, and by no
means affect the universal properties of the model. For practical purposes, therefore,
Eq.~(\ref{eq:1}) is sufficiently general.

\subsection{Decoupling of the model Hamiltonian}
\label{sec:2}
It is straightforward to construct linear combinations of the operators $\cn{k\alpha}$ ($\alpha=L,r$)
that are decoupled from the quantum dot \cite{PuG04:R513}. To this end we define the orthonormal Fermi
operators 
  \begin{align}\label{eq:2}
    \an{k} &= \dfrac{V\sub{L}\cn{kL} +V\sub{R}\cn{kR}}{V}\\
    \bn{k} &= \dfrac{V\sub{R}\cn{kL} -V\sub{L}\cn{kR}}{V},\label{eq:3}
  \end{align}

where
\begin{align}
  \label{eq:4}
  V \equiv \sqrt{V_{L}^2+V_{R}^2}.
\end{align}

Projected on the basis of $\an{k}$'s and $\bn{k}$'s, the Hamiltonian~(\ref{eq:1}) reduces to the
form
\begin{align}
  \label{eq:5}
  H\sub{A} = H + \bar H,
\end{align}
where
\begin{align}
  \label{eq:6}
  \bar H = \sum\sub{k}\epsilon\sub{k}\bd{k}\bn{k}+\dfrac{\vw}{N}\sum\sub{k,q}\bd{k}\bn{q},
\end{align}
and
\begin{align}\label{eq:7}
  H = H_d + \sum\sub{k}\epsilon\sub{k}\ad{k}\an{k}+
  \vw\fd{0}\fn{0} +V(\cd{d}\fn{0}+\hc),
\end{align}
with the shorthand
\begin{align}
  \label{eq:8}
  \fn{0} = \dfrac1{\sqrt N}\sum\sub{k}\an{k}.
\end{align}

The right-hand side of Eq.~(\ref{eq:7}) is the standard expression for the single-impurity,
single-band Anderson Hamiltonian \cite{An61:41}. The second band, defined by Eq.~(\ref{eq:6}), is
decoupled from the quantum dot and can be disregarded, for nearly all applications. Exceptions are
the transport properties, to which the $\bn{k}$'s contribute. To
compute the zero-bias electrical conductance, for example, one must apply an infinitesimal bias
\begin{align}
  \label{eq:9}
H\sub{\mu} = \Delta\mu \sum\sub{k}\Big(\cd{kR}\cn{R}-\cd{kL}\cn{kL}\Big),
\end{align}
between the $L$ and the $R$ leads.

Projection of Eq.~(\ref{eq:9}) upon the basis defined by Eqs.~(\ref{eq:2}) shows that $H\sub{\mu}$
couples the $\an{k}$'s to the $\bn{k}$'s. Likewise, the current operator $\hat I=d\hat q\sub{R}/t$,
where $\hat q\sub{R}$ is the electrical charge in lead $R$, couples the $\an{k}$'s to the
$\bn{k}$'s. Standard linear response theory links the conductance $G$ to the commutator
between the bias Hamiltonian $H\sub{mu}$ to the current $\hat I$. 

This considered, one can follow the algebraic
manipulations in appendix~C of Ref.~\onlinecite{YSO2009:235317} to show that
\begin{align}
  \label{eq:10}
  G(T) =   \kappa\pi\gtwo
  \Gamma\sub{\vw} \int\left(-\frac{\partial f}{\partial\epsilon}\right)
  \rho_d(\epsilon)\,\text{d}\epsilon,
\end{align}
where $\rho\sub{d}(\epsilon)$ is the $\cn{d}$-level spectral density, $f(\epsilon)$ is the Fermi
function,
\begin{align}\label{eq:11}
  \gtwo \equiv \dfrac{2\mathrm{e}^{2}}{hc}
\end{align}
is the quantum conductance with two spin channels,
\begin{align}\label{eq:12}
  \Gamma\sub{\vw} = \dfrac{\pi\rho V\super{2}}{1+\pi\super{2}\rho\super{2}\vw\super{2}},
\end{align}
and
\begin{align}\label{eq:13}
    \kappa = \frac{V\sub{L}V\sub{R}}{V\super{2}}.
\end{align}
The asymmetry index $\kappa$ is a dimensionless factor that modulates the conductance.  The
modulus is unitary for symmetric couplings, $V\sub{L}=V\sub{R}$, and shrinks as the coupling
asymmetry grows. To simplify the following theoretical analysis, we define the \emph{reduced
 conductance}
\begin{align}
  \label{eq:14}
  \bar G \equiv \dfrac{G}{\kappa},
\end{align}
so that Eq.~\eqref{eq:10} reads
\begin{align}
  \label{eq:15}
  \bar G(T) =   \pi\gtwo
  \Gamma\sub{\vw} \int\left(-\frac{\partial f}{\partial\epsilon}\right)
  \rho_d(\epsilon)\,\text{d}\epsilon.
\end{align}

\section{Characteristic energies and fixed points}
\label{sec:3}
The spectral density $\rho\sub{d}$ is a function of energy and temperature. Since the
Hamiltonian~(\ref{eq:6}) is decoupled from the dot, we only have to diagonalize $H$ to compute
$\rho\sub{d}$ and determine the conductance from Eq.~(\ref{eq:15}). The computation is simple in
special regimes, defined by the characteristic energies of the Hamiltonian. 

In the absence of the potential $\vw$, the coupling to the leads broadens the dot level to the width
\begin{align}
  \label{eq:16}
  \Gamma = \pi\rho V\super{2}.
\end{align}
The potential $\vw$ reduces the broadening, as indicated by Eq.~(\ref{eq:12}).

If the width $\Gamma$ were zero, the dot occupation $n\sub{d}$ would be conserved. The dynamics of the
device would then be controlled by the eigenvalues $E_{d}^{\ell}$ ($\ell=0,1,2$) of
$H\sub{d}$. The $d\super{0}$ eigenstate would have energy $E_{d}^{0}=0$, the $d^{1}_{\uparrow}$ and
$d^{1}_{\downarrow}$ eigenstates (where the subscript indicates the $S\sub{z}$ eigenvalue) would have energy
 $E_{d}^{1}= V\sub{G}$, and the $d\super{2}$ eigenstate would have energy $E_{d}^{2}= 2V\sub{G}+U$.

We are centrally interested in the gate-voltage range making $E_{d}^{1}$ smaller than $E_{d}^{0}$
and $E_{d}^{2}$, \ie\ in the range $0> V\sub{G} >-U$. In this interval, the dot acquires a
magnetic moment $\mu\sub{B}$. The interval is limited by the two \emph{charge-degeneracy points},
associated with voltages $V_{G}^{0\to1}=0$ and $V_G^{1\to2}=-U$. At the middle of the interval is
the \emph{symmetric point}, attained when the gate voltage is $V^{1}_{G}=-U/2$.

With no coupling, conduction would be impossible. With small coupling $\Gamma\sub{\vw}\ll
|V\sub{G}|$, $\Gamma\sub{\vw} \ll U$, charge transport is barred by an
energy barrier 
$\Delta E\sub{c}= \min\{|V\sub{G}|, U+V\sub{G}\}$, 
except within a gate-voltage range of width $\Gamma\sub{\vw}$ of either
charge-degeneracy point. The barrier $\Delta E\sub{c}$ is known as the \emph{Coulomb blockade}.

At moderately high temperatures, such that thermal energy $k\sub{B}T$
lies in the interval $\Delta E\sub{c} \gg k\sub{B}T \gg T\sub{K}$, the
width $\Gamma$ can be disregarded, the dot occupation is approximately
conserved, and the Coulomb blockade controls the physics of
conduction\textemdash the Coulomb blockade regime.  Assuming that the
width $D$ of the conduction bands exceeds $U$, we can see that the
thermal energy is incommensurate with the other energy scales of the
problem. Physically, the model Hamiltonian is then approximately
equivalent to the \emph{local-moment fixed-point} Hamiltonian
$H_{LM}^{*}$ obtained by letting $D,U\to\infty$,
$V\sub{G}\to -\infty$, and $\Gamma\to 0$ in Eq.~(\ref{eq:7}).

The local-moment fixed-point fixed-point Hamiltonian is equivalent
to a dot with unitary occupation and magnetic moment $\mu\sub{B}$
decoupled from conduction band of non-interacting electrons. The
Hamiltonian reads
\begin{align}
  \label{eq:17}
H_{LM}^{*}= \sum\sub{k}\epsilon\sub{k}\ad{k}\an{k} + \vw\fd{0}\fn{0}.
\end{align}
where the superscript reminds us that the fixed-point Hamiltonian is devoid of characteristic
energies. 

$H_{LM}^{*}$ is an idealized Hamiltonian whose many-body spectrum is approximately equal
to the energy spectrum of the Hamiltonian $H$ in the range
$\Delta E\sub{c} \gg \epsilon\gg \Gamma\sub{\vw}$. The dot level makes no contribution to
the \rhs. Still, the dot level has a spin-$1/2$ degree of freedom, which we will denote
$\vec S$.

The conduction band is phase shifted by the potential $\vw$, \ie\ each single-particle eigenstate
acquires a phase shift $\delta\sub{\vw}$, given by the expression
\begin{align}
  \label{eq:18}
\tan\delta\sub{\vw} = -\pi\rho \vw.
\end{align}
Depending on $\vw$, the phase shift can take any value in the interval
$-\pi/2\le\delta/2\le\pi/2$. The LM fixed-point Hamiltonian can be visualized as a point with phase
shift $\delta$ along a line running from $-\pi/2$ to $\pi/2$.

\subsection{Kondo Hamiltonian}
\label{sec:4}

The fixed point is an idealization. In practice, neither $U$, nor $|V\sub{G}|$ are infinite. Even at
(moderately) high temperatures, the Hamiltonian $H$ is not exactly the fixed-point Hamiltonian. The
high-energy many-body spectrum of $H_{LM}^{*}$ is only an approximation to the spectrum of $H$,
because the finite Coulomb barriers allow virtual excitations to the $d^{0}$ and $d^{2}$ dot
states. 

The virtual excitations induce an antiferromagnetic coupling between the dot magnetic moment
and the magnetic moments of the conduction electrons \cite{SW66:491}. A more precise representation
of the high-energy spectrum of $H$ comes therefore from the equation
\begin{align}
  \label{eq:19}
  H\sub{K} = \sum\sub{k}\epsilon\sub{k}\ad{k}\an{k} + \tilde\vw\fd{0}\fn{0} 
  + J\vec{S}\cdot\sum\sub{\mu,\nu}\vec{\sigma}\sub{\mu\nu}\fd{0\mu}\fn{0\nu},
\end{align}
where the components of $\vec\sigma$ are the Pauli matrices, and the coefficientes of the second
and third terms on the \rhs\ are given by the Schrieffer-Wolff expressions \cite{SW66:491}
\begin{align}
  \label{eq:20}
\rho \tilde\vw = \rho\vw + \dfrac{\Gamma}{V\sub{G}} + \dfrac{\Gamma}{V\sub{G}+U} 
\end{align}
and
\begin{align}\label{eq:21}
  \rho J = \dfrac{\Gamma}{|V\sub{G}|} + \dfrac{\Gamma}{V\sub{G}+U}.
\end{align}

Equation~(\ref{eq:19}) defines the Kondo Hamiltonian. For thermal energies that are small in the
scale of the Coulomb blockade, the spectra of $H$ and $H\sub{K}$ are approximately congruent. The
{\rhs}s of Eqs.~\eqref{eq:20}~and \eqref{eq:21} become very large in absolute value near the charge
degeneracy points $V\sub{G}=0$ and $V\sub{G}=-U$. 

Near the symmetric point, by contrast, as long as $\Gamma \ll U$ there is a range of gate voltages
such that $\Gamma\ll|V\sub{G}|$ and $\Gamma\ll U+V\sub{G}$. That
gate-voltage range makes $\rho J \ll 1$ and places the device in the \emph{Kondo regime}. 

The symmetric point lies at the middle of the Kondo regime. At the symmetric point,
the second and third terms on the \rhs\ of Eq.~\eqref{eq:20} cancel each other, and the phase shift
equals $\delta\sub{W}$. Elsewhere within the Kondo regime, the phase shift is given by the equality
\begin{align}
  \label{eq:22}
  \tan\delta\sub0 = -\pi\Bigg(\rho\vw + \dfrac{\Gamma}{V\sub{G}} + \dfrac{\Gamma}{V\sub{G}+U}\Bigg).
\end{align}

Physically, the phase shift is associated with the screening charge that forms in the vicinity of
the $\fn{0}$ orbital in response to the potential $\vw$ and to the coupling to the quantum dot. 
\subsection{Frozen-level fixed point}
\label{sec:5}

If the device is cooled in the Kondo regime, at sufficiently low temperatures the antiferromagnetic
interaction between the conduction elctrons and the dot spin will induce the Kondo cloud. At
temperatures well below the Kondo temperature, the dot spin will lock into a singlet with the
conduction-electron spins, which will freeze the dot-spin degree of freedom.

As a result, at low thermal energies, with $T\ll T\sub{K}$, the spectrum of $H$ approaches that of
the Hamiltonian obtained from Eq.~(\ref{eq:19}) when we let $J\to\infty$. The $\fn{0}$ orbital then
forms a singlet with the dot spin variable, and the Hamiltonian becomes equivalent to the quadratic
form
\begin{align}
  \label{eq:25}
  H_{FL}^{*}=\sum\sub{k}\bar\epsilon\sub{k} \bad{k}\ban{k}+\tilde\vw\sum\sub{k,q}\bad{k}\ban{q},
\end{align}
where the set of the conduction states $\ban{k}$ and the localized orbital $\fn{0}$ form
an orthonormal basis that is complete relative to the original conduction states
$\an{k}$. The subscript on the left-hand side reminds us that the dot level is frozen, and
the superscript, that $H_{FL}^{*}$ is devoid of characteristic energies.

To be orthogonal to $\fn{0}$ the new conduction states $\ban{k}$ must deplete the region of the
leads next to the quantum dot. They must therefore be phase-shifted by $\pi/2$ relative to the
$\an{k}$. It follows that the conduction energies $\bar\epsilon\sub{k}$ are shifted relative to
the $\epsilon\sub{k}$:
\begin{align}
  \label{eq:26}
  \rho\bar\epsilon\sub{k} = \rho\epsilon\sub{k}-\dfrac{1}{2},
\end{align}
and that the FL fixed-point phase shift is
\begin{align}
  \label{eq:72}
  \delta = \dfrac{\pi}2+\delta\sub{0},
\end{align}
where $\delta\sub{0}$ is the LM fixed-point phase shift.

\subsection{Fixed-point conductances}
\label{sec:19}

To determine the conductance from Eq.~\eqref{eq:15}, we must compute the spectral density
$\rho\sub{d}(\epsilon)$. An exact expression relates $\rho\sub{d}(\epsilon)$ to the spectral
densities of the linear combinations $\sum_{k}\an{k}$ and $\sum_{k}\epsilon\sub{k}\an{k}$ of the
conduction operators $\cn{k}$ \cite{YSO2009:235317,Pinto20141299}. As the model Hamiltonian
approaches a fixed point, the latter two spectral densities can be computed from the eigenvalues and
eigenstates of the fixed-point Hamiltonian. The diagonalization of the fixed-point, single-particle
Hamiltonians $H_{LM}^{*}$ and $H_{FL}^{*}$ is straightforward. It is therefore a simple matter to obtain
the fixed-point spectral densities \cite{YSO2009:235317}
\begin{align}
  \label{eq:32}
  \rho_{d}^{*}= \dfrac{1}{\pi\Gamma\sub{\vw}}\sin^{2}(\delta\sub{*}-\delta\sub{\vw})
  \qquad(T\gg T\sub{K}\mbox{\ or\ } T\ll T\sub{K}),
\end{align}
where $\delta\sub{*}$ denotes the fixed-point phase shift.

For $\vw=0$, the phase shift $\delta\sub{\vw}$ vanishes and we recover Langreth's expression for the
low-energy spectral density \cite{La66:516}. Equation~\eqref{eq:32} is not restricted to low
energies. However, since the LM and the FL fixed points have distinct phase-shifts, the
spectral densities at high and at low energies are different.

Substitution of the fixed-point results in Eq.~\eqref{eq:15} now yields the following expression
for the fixed-point conductances:
\begin{align}
  \label{eq:33}
  \bar G\super{*}  = \gtwo \sin^{2}(\delta\sub{*}-\delta\sub{\vw}).
\end{align}

\subsection{Thermal dependence of the conductance}
\label{sec:20}

If $\vw=0$, the zero-bias conductance at the symmetric point is a universal function of the
temperature scaled by the Kondo temperature \cite{CHZ94.19,BCP08:395}:
\begin{align}
  \label{eq:34}
  \bar G(T) = \gtwo\guniv(T/T\sub{K}).
\end{align}

\begin{figure}[htb!]
  \centering
  \includegraphics[width=0.95\columnwidth]{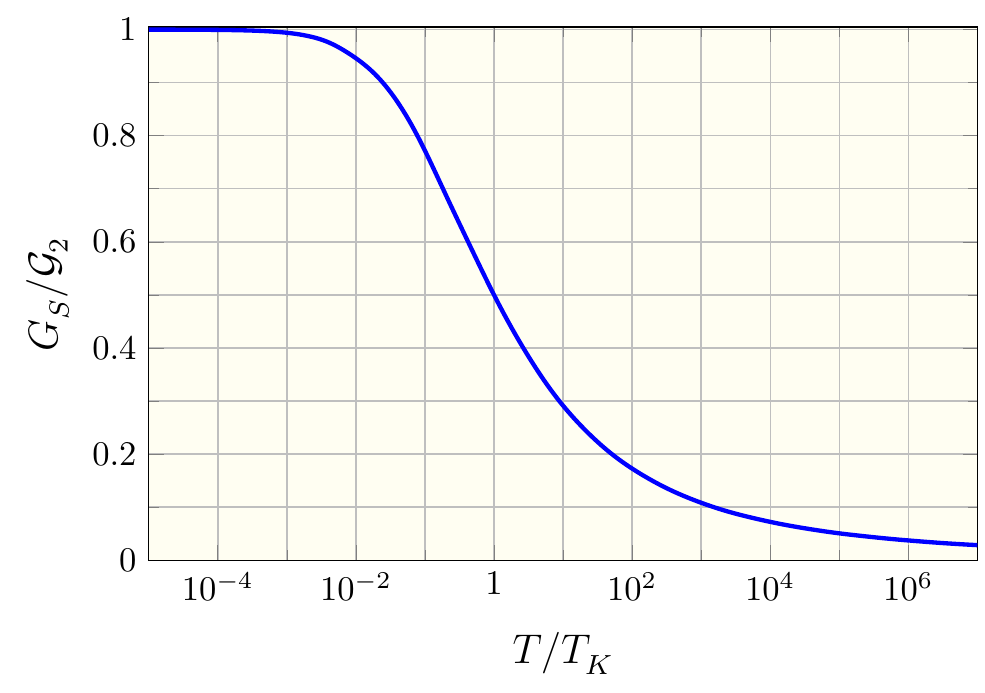}
  \caption[Guniv]{Universal function relating the SET conductance at the symmetric point
    $V\sub{G}=-U/2$ to the temperature scaled by the Kondo temperature.}
  \label{fig:2}
\end{figure}

Figure~\ref{fig:2} displays the universal function $\guniv$ as a function of the ratio $T/T\sub{K}$.
The condition $\guniv(T=T\sub{K})=\gtwo/2$ defines the Kondo temperature. At high (low) temperatures, the
Hamiltonian is close to the LM (FL) fixed point, and the conductance, close to zero
($\gtwo$). Physically, the coupling $J$ between the dot and the conduction-electron spins is so weak
that the dot moment is virtually decoupled from the leads. At the symmetric point the Coulomb
blockade imposes the energy barrier $\Delta E\sub{c}= U/2$, much larger than the thermal
energy. Conduction across the device is virtually impossible. 

As the temperature is lowered, the Kondo cloud starts forming. As $T$ drops past $T\sub{K}$ the
electrons within the cloud bind into a singlet with the dot electron. The binding is so tight that
it allows ballistic transport.

\subsubsection{Particle-hole symmetry}
\label{sec:21}
The symmetric point is special. With $\vw=0$ and $U=-V/2$, the Hamiltonian~\eqref{eq:7} remains
invariant under the particle-hole transformation
\begin{align}
  \cn{d}&\to -\cd{d};\ncr
  \cn{k}&\to \cd{q}.\label{eq:35}
\end{align}
Here, the momenta $k$ and $q$ are symmetric: given $k$, one chooses $q$ such that
$\epsilon\sub{q}=-\epsilon\sub{k}$.

The particle-hole transformation inverts the sign of the phase shift $\delta$. Consequently, the
fixed-point phase shifts of the symmetric Hamiltonian can only be $\delta=0$ or $\delta=\pi/2$. At
the high-temperature (LM) fixed point the phase shift vanishes; at the low-temperature (FL) fixed
point $\delta=\pi/2$. It follows from Eq.~\eqref{eq:33} that, at the symmetric point, $\bar G^{*}_{LM}=0$
and $\bar G^{*}_{FL}=\pi/2$, as indicated by the high- and low-temperature limits in Fig.~\ref{fig:2}.

\subsubsection{Linear mapping}
\label{sec:22}
An applied potential $\vw$, or deviation from the condition $V\sub{G}=-U/2$, breaks
particle-hole symmetry.  Depending on the model parameters, the fixed-points phase shifts
can now take any values in the $[-\pi/2,\pi/2]$ interval. The LM (FL) conductance will no
longer be zero ($\gtwo$). Clearly, $\bar G(T/T\sub{K})$ cannot follow the plot in
Fig.~\ref{fig:2}, nor can it be proportional to $\guniv(T/T\sub{K})$.

Instead, the conductance maps linearly onto the universal function
\cite{SYO09:67006,YSO2009:235317}:
\begin{align}
  \label{eq:36}
  \bar G\Big(\dfrac{T}{T\sub{K}}\Big) =\alpha \guniv\Big(\dfrac{T}{T\sub{K}}\Big) + \beta.
\end{align}

To determine the linear coefficients $\alpha$ and $\beta$, we go back to Eq.~\eqref{eq:33}.  At the
FL fixed point, the phase shift is $\delta$, and the universal conductance reaches
$\gtwo$ as $T\to0$. Equation~\eqref{eq:36} then reads
\begin{align}
  \label{eq:27}
    \gtwo \sin^{2}(\delta-\delta\sub{\vw}) = \alpha \gtwo + \beta.
\end{align}

At the LM fixed point, the phase shift is $\delta\sub{0}=\delta-\pi/2$, and the universal
conductance vanishes in the large $T$ limit. Equation~\eqref{eq:36} therefore reads
\begin{align}
  \label{eq:28}
  \gtwo\cos\super{2}(\delta-\delta\sub{\vw}) = \beta.
\end{align}

Substitution on the \rhs\ of Eq.~\eqref{eq:27} determines $\alpha$ and brings Eq.~\eqref{eq:36}
to the explicit form
\begin{align}
  \label{eq:37}
  \bar G\Big(\dfrac{T}{T\sub{K}}\Big) -\dfrac{\gtwo}{2}=
  \left(\dfrac{\gtwo}{2}-\guniv\Big(\dfrac{T}{T\sub{K}}\Big)
  \right)\cos2(\delta-\delta\sub{\vw}).
\end{align}

At the symmetric point, with $\vw=0$ the FL phase shift is $\delta=\pi/2$, and
Eq.~\eqref{eq:37} reduces to Eq.~\eqref{eq:34}. Elsewhere in the parametrical space of the
model, the trigonometric function on the \rhs\ is larger than $-1$. The low-temperature
(high-temperature) conductance is then positive (smaller than $\gtwo$). At the Kondo
temperature, the conductance is always $\gtwo/2$, but the difference between $\bar
G_{FL}^{*}$  and $\bar G_{LM}^{*}$ may be significantly smaller than $\gtwo$.

\subsubsection{Phase shifts and occupation}
\label{sec:23}
The Friedel sum rule relates the fixed-point phase shifts to the dot occupation. At the FL fixed
point, the occupation $n\sub{d}$ is proportional to the phase shift induced by the coupling to the
leads. If there were no coupling, the phase shift would be $\delta\sub{\vw}$. The induced shift is
therefore $\delta\sub{*}-\delta\sub{\vw}$. According to the Friedel sum rule \cite{La66:516}, then,
\begin{align}
  \label{eq:38}
  n\sub{d} = \dfrac{2(\delta\sub{*}-\delta\sub{\vw})}{\pi}.
\end{align}

Equation~\eqref{eq:37} can now be rewritten in the form
\begin{align}
  \label{eq:71}
  \bar G\Big(\dfrac{T}{T\sub{K}}\Big) -\dfrac{\gtwo}{2}=
  \left(\dfrac{\gtwo}{2}-\guniv\Big(\dfrac{T}{T\sub{K}}\Big)
  \right)\cos(\pi n\sub{d}),
\end{align}
which shows that the thermal dependence of the conductance is parametrized by the ground-state
expectation value for the dot occupancy.

\section{Comparison with experiment}
\label{sec:24}
In the Kondo regime, Eq.~\eqref{eq:71} is exact. It is instructive to compare it with the
experimental data reported in Ref.~\onlinecite{GRP+08:246601}. Equation~\eqref{eq:14}, which relates
the experimental conductance $G(T)$ to the reduced conductance $\bar G(T)$, yields the
following expressions for the high- and low-temperature limits of the experimental conductance:
\begin{align}
  \label{eq:29}
  G\sub{LM} = \kappa\gtwo\cos\super{2}\Big(\dfrac{\pi n\sub{d}}2\Big),
\end{align}
and 
\begin{align}
  \label{eq:30}
  G\sub{FL} = \kappa\gtwo\sin\super{2}\Big(\dfrac{\pi n\sub{d}}2\Big).
\end{align}

At intermediate temperatures, the experimental conductance maps linearly onto the universal
function. From Eq.~\eqref{eq:71} it follows that
\begin{align}
  \label{eq:31}
  G\Big(\dfrac{T}{T\sub{K}}\Big) -\kappa\dfrac{\gtwo}{2}=
  \left(\dfrac{\gtwo}{2}-\guniv\Big(\dfrac{T}{T\sub{K}}\Big)
  \right)\kappa\cos(\pi n\sub{d}).
\end{align}

\subsection{Experimental data}
\label{sec:6}
Grobis \etal\ \cite{GRP+08:246601} have measured the conductance of a single-electron transistor as a
function of temperature, gate voltage, and bias voltage. We focus on their zero-bias results. To scan a
Kondo plateau, the authors have accurately measured $G$ on a $V\sub{G}\times T$ grid 
comprising 34 uniformly spaced gate-voltages, ranging from \SI{-212.5}{mV} to \SI{-196}{mV}, and 17
temperatures, ranging from \SI{13}{mK} to \SI{205}{mK}. Figure 1(c) in
Ref.~\onlinecite{GRP+08:246601} overviews the resulting data. At fixed gate-voltage, the conductance
rises as the sample is cooled, from approximately $0.5\gtwo$ at $T=\SI{205}{mK}$ to approximately
$0.85\gtwo$ at $T=\SI{13}{mK}$. The rise is steeper at the middle of the plateau, around
$V\sub{G}=\SI{-205}{mV}$.

\begin{figure}[htb!]
  \centering
  \includegraphics[width=0.85\columnwidth]{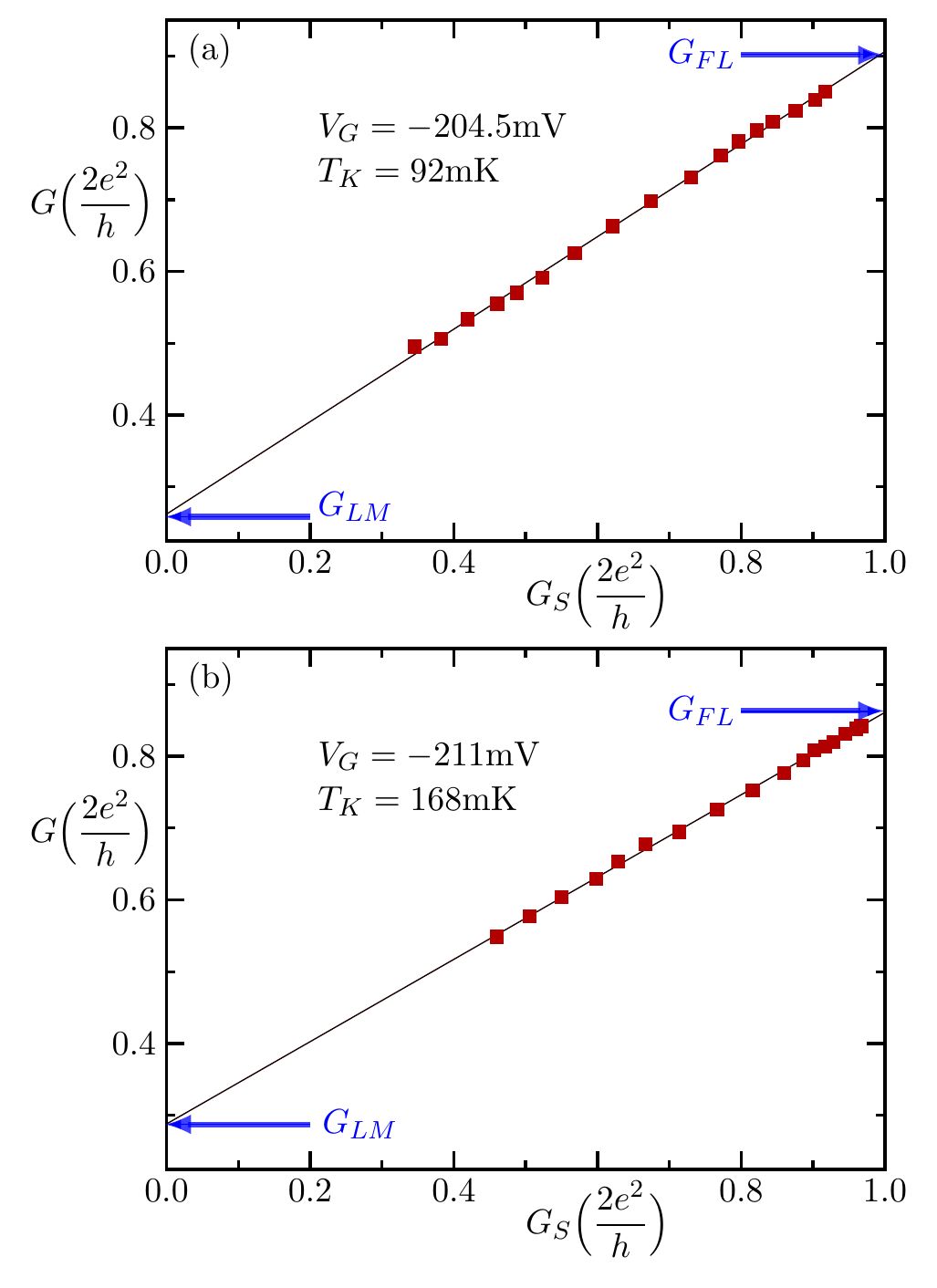}
  \caption[all VGs]{(Color online). Conductance $G(T/T_K)$ as a function of the
    universal function $\guniv(T/T_K)$. The filled squares in
    (a) and (b) show the conductances measured in
    Ref.~\onlinecite{GRP+08:246601} for the indicated gate voltages at
    17 temperatures ranging from $T=\SI{12.5}{mK}$ (point closest to
    the upper right corner) to $T=\SI{203}{mK}$ (closest to the
    bottom left corner). In each plot, the indicated Kondo temperature
    optimizes the linear regression of $G(T/T\sub{K})$ vs.~$G\sub{SET}(T/T\sub{K})$,
    depicted by a solid line. The horizontal arrows labeled $G_{FL}$
    and $G_{LM}$ point to the low- and high-temperature limits of the
    conductance, respectively.
   }
  \label{fig:3}
\end{figure}

Qualitatively, we can see that Eq.~\eqref{eq:31} agrees with these features of the
data. In fact, the agreement is quantitative, as illustrated by Fig.~\ref{fig:3}. Each panel plots
the measured conductance $G$ as a function of the universal conductance $\guniv$ for the depicted
gate voltage. As long as the temperature is scaled by the Kondo temperature, we expect the
relation between the two conductances to be linear.

Since $T\sub{K}$ is unknown, we proceed by trial and error. The experimental temperatures are scaled
by a trial Kondo temperature, and linear regression determines the optimum coefficient
$\kappa\cos(\pi n\sub{d})$ and intercept $\kappa\gtwo/2$ fitting $G(T/T\sub{K})$ to
$\guniv(T/T\sub{K})$. If the linear correlation coefficient is sufficiently close to unity, we have
found the Kondo temperature. Otherwise, we turn to Newton's method for a better estimate of
$T\sub{K}$, and repeat the procedure. Convergence yields the Kondo temperature and the coefficients
of the linear fit.

This procedure was applied to the thermal dependence of the conductance at each gate voltage in the
experimental grid. In each case, the agreement was comparable to the fits in
Fig.~\ref{fig:3}. 

As the plots in Fig.~\ref{fig:3} show, the straight lines can be extrapolated to the $\guniv\to \gtwo$
($\guniv\to0$) limit to yield the FL (LM) fixed-point conductance $G\sub{FL}$
($G\sub{LM}$). Linear regression therefore determines the high- and the low-temperature limits of
the conductance, which are inaccessible in the laboratory.

\begin{figure}[h!]
  \centering
\includegraphics[width=0.85\columnwidth]{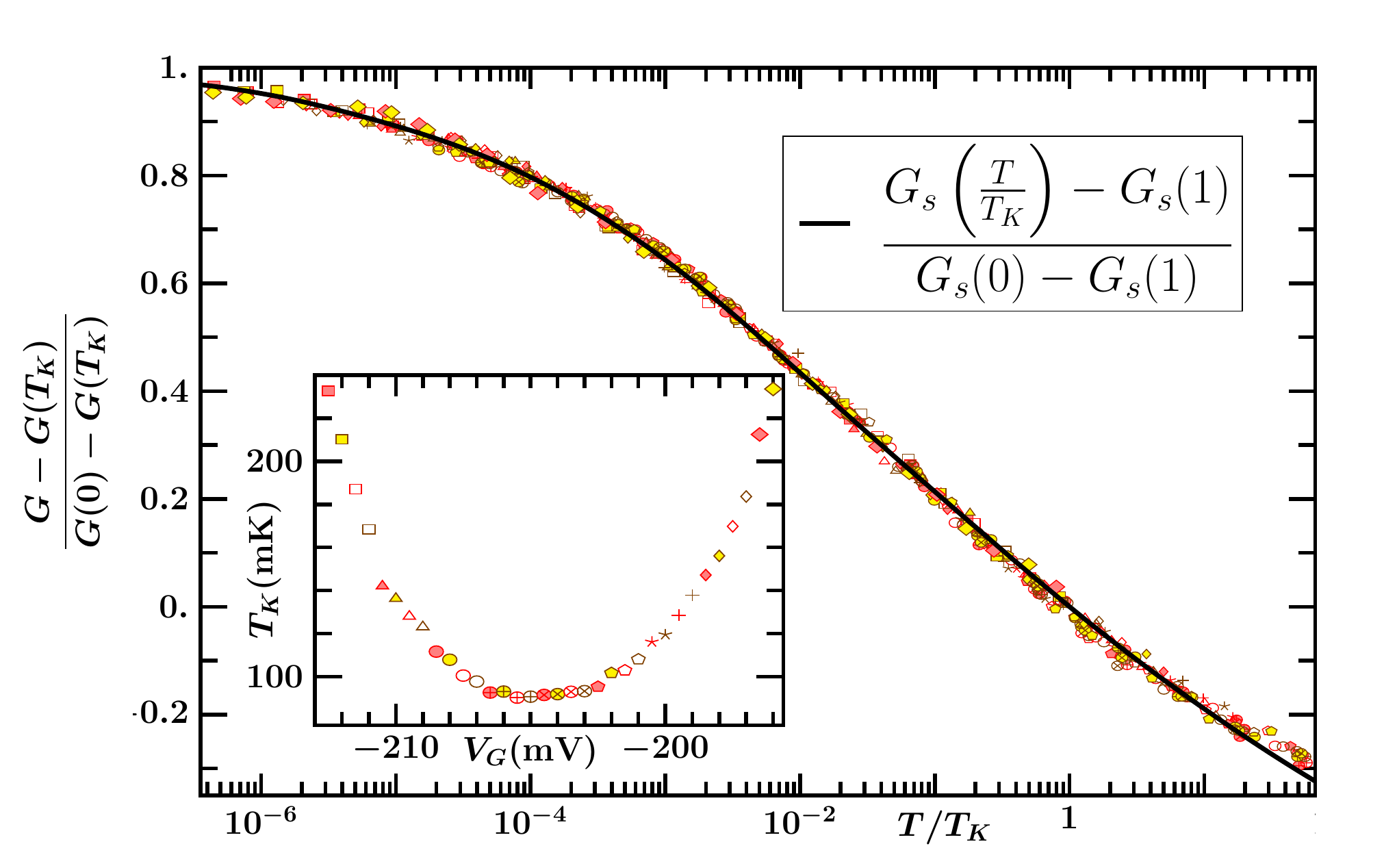}
\caption[scaled]{(Color online) Scaled zero-bias conductance $G(T)$ (symbols)
  and universal function $G_S(T/T_K)$ (solid line), as functions of
  $T/T_K$. Each symbol represents the former for one of the 34 gate
  voltages $V_G$ in \grp. In the inset, the same symbol shows the
  corresponding $T_K$.}
\label{fig:4}
\end{figure}

Figure~\ref{fig:4} shows all $34\times17\equiv578$ experimental conductances, measured
from $G(T\sub{K})=\gtwo/2$, scaled by the difference $G(0)-G(T\sub{K})$ between the
extrapolated low-temperature conductance and the Kondo-temperature conductance, as a
function of the temperature scaled by $T\sub{K}$. To identify the gate-voltage at which
each conductance was measured, the inset shows the 34 Kondo temperatures as a function of
$V\sub{G}$. The near congruence between the symbols and the solid line representing the
universal function scaled in the same fashion offers a measure of the overall harmony
between the measurements and the expected universal behavior in the Kondo regime.

\begin{figure}[htb!]
  \centering
  \includegraphics[width=0.85\columnwidth]{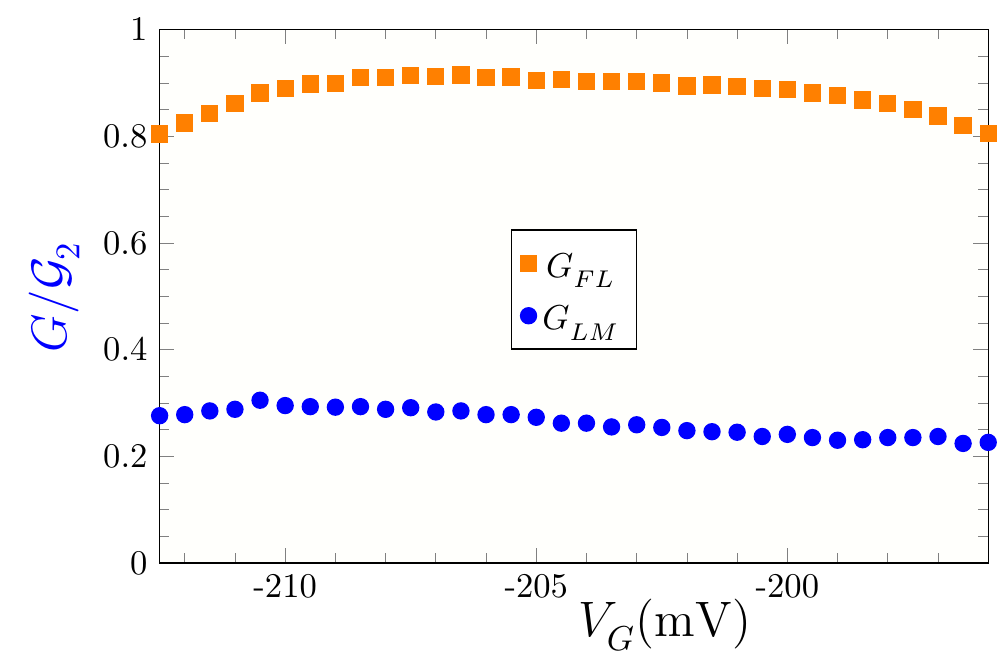}
  \caption[fixed-point]{(Color online) Fixed-point conductances $G\sub{LM}$ and $G\sub{FL}$ as
    functions of the gate voltage $V\sub{G}$. At each gate voltage, the LM and FL fixed-point
    conductances result from extrapolating the $G(T/T\sub{K})$ \emph{vs.}~$\guniv(T/T\sub{K})$ plots
    to $\guniv=0$ and $\guniv=1$, respectively, as illustrated by each panel in Fig.~\ref{fig:4}.}
  \label{fig:5}
\end{figure}

Figure~\ref{fig:5} shows the resulting estimates of $G\sub{FL}$ and $G\sub{LM}$ as
functions of the gate voltage. At each gate-voltage, the limit conductances determine
the ground-state expectation value $n\sub{d}$ of the dot occupancy and the asymmetry index
$\kappa$. To obtain $n\sub{d}$ from Eqs.~\eqref{eq:29}~and \eqref{eq:30}, we compute the
ratio
\begin{align}
  \label{eq:39}
  \dfrac{G\sub{FL}}{G\sub{LM}} = \tan\super2\Big(\dfrac{\pi n\sub{d}}2\Big).
 \end{align}
To obtain the asymmetry index we compute the sum
\begin{align}
  \label{eq:40}
   \dfrac{G\sub{FL}}{\guniv}+\dfrac{G\sub{LM}}{\guniv} = \kappa.
\end{align}

\begin{figure}[htb!]
  \centering
  \includegraphics[width=0.85\columnwidth]{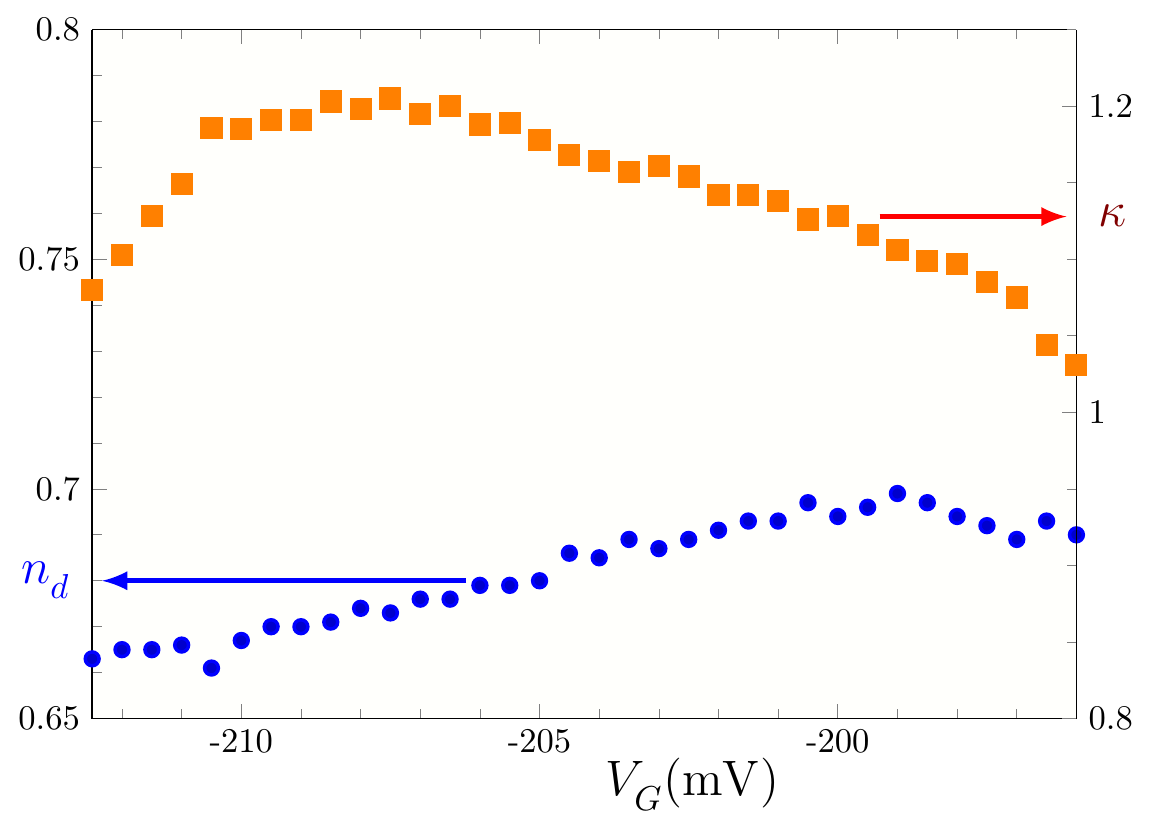}
  \caption[asymmetry]{(Color online) Coupling asymmetry index $\kappa$ and ground-state dot-level
    occupation $n\sub{d}$ as functions of the gate voltage $V\sub{G}$. At each gate voltage,
    Eq.~\eqref{eq:40} determines $\kappa$, while Eq.~\eqref{eq:39} determies $n\sub{d}$.}
  \label{fig:6}
\end{figure}

The resulting dot occupancy $n\sub{d}$ and asymmetry index $\kappa$ are depicted in Fig.~\ref{fig:6},
as functions of the gate voltage $V\sub{G}$. Both plots have unexpected features. In the Kondo regime,
the ground-state dot occupancy should be close to unity. Instead, the blue dots in the figure span an
interval ranging from $n\sub{d}=0.66$ to $n\sub{d}=0.70$. The asymmetry index, expected to be
a constant $\kappa\le1$, varies between $\kappa=1.03$ and $\kappa=1.20$.

We conclude that the Hamiltonian~\eqref{eq:1} cannot describe the device in
Ref.~\cite{GRP+08:246601} quantitatively. Given the simplicity of the model, which neglects
electron-electron interactions within the leads, considers a single, structureless, half-filled
conduction bands, adopts a single level to represent the quantum dot, and gives no attention to the
spatial dependence of the potentials applied to the leads or to the momentum dependence of the
couplings between the dot and the leads, the conclusion seems hardly surprising.

More puzzling is the contrast between the deviations of Fig.~\ref{fig:6} from the expected
behavior and the much superior agreements in Figs.~\ref{fig:3}~and \ref{fig:4}. The
puzzle, however, is easily solved. Recall that Eq.~\eqref{eq:35}, besides exact in the
Kondo regime, is universal. Other properties of the Anderson model, such as the
mathematical connection between the physical features of the device and the LM
Hamiltonian, are nonuniversal. The minimalist, inaccurate description of the physical
features is responsible for the deviations in Fig.~\ref{fig:6}, while universality
protects the fits in Figs.~\ref{fig:3}~and \ref{fig:4} from such inaccuracies.

This underscores the call for \emph{ab initio} treatments of the device. Consider, for
definiteness, the experimental data by Grobis et al.~\cite{GRP+08:246601}. While static
DFT approaches may be unable to deal with the crossover from the LM to FL fixed points,
even local approximations to the exchange-correlation functional should be sufficient to
describe the LM fixed point and determine the LM conductance. Comparison with the solid
blue circles in Fig.~\ref{fig:5} would then test our understanding of the physics
underlying the experimental data.

To examine the same argument from a different perspective, consider a local-density
description of the LM fixed point associated with the model
Hamiltonian~\eqref{eq:7}. $H_{LM}^{*}$ is given by Eq.~\eqref{eq:17}. Since the model
conduction electrons are noninteracting, the exchange-correlation potential vanishes
inside the leads. The Kohn-Sham eigenstates are the single-particle eigenstates of
$H_{LM}^{*}$. DFT therefore yields the phase shift $\delta=\delta\sub{W}$. From
Eqs.~\eqref{eq:14}~and \eqref{eq:33}, the conductance is predicted to vanish at high
temperatures, in the Kondo regime. At the symmetric point, this agrees with the plot in
Fig.~\ref{fig:2}. The prediction nonetheless disagrees, conspicuously, with the blue solid
circles in Fig.~\ref{fig:5}.

DFT cannot be blamed for the disagreement. True, the local-density approximation neglects the
antiferromagnetic interaction with the dot spin and hence misses the contribution from the last term
on the \rhs\ of Eq.~\eqref{eq:19} and the contribution to $\tilde\vw$ from the last two terms on the
\rhs\ of Eq.~\eqref{eq:20}. In the Kondo regime, however, those terms are small. They cannot account
for the substantial conductances represented by the blue solid circles in Fig.~\ref{fig:5}.

The discrepancy is due to the shortcomings of the model, not to the limitations of the
local-density approximation for the exchange-correlation functional. To strengthen the
argument, we substitute $\tilde\vw$ [given by Eq.~\eqref{eq:20}] for $\vw$ on the \rhs\ of
Eq.~\eqref{eq:17}. A more accurate approximation results, which takes the spin-independent
phase shift induced by dot moment into account. Notwithstanding the improvement, the
resulting conductance is still zero at the symmetric point, which corresponds to
$V\sub{G}\approx\SI{-205}{mV}$ in the experimental setup \cite{GRP+08:246601}.

The model fails to account for the relatively large LM conductances resultant
from the extrapolations of the experimental data. A mored detailed description of the experimental
device, based on \emph{ab inition} computations, is necessary to describe the LM fixed point.

\section{Summary}
\label{sec:7}

The single-electron transistor poses a concrete challenge to DFT. Recent progress, backed
by improved local approximations for the exchange-correlation functional especially
designed to yield the correct density derivative, have yielded accurate descriptions of
the ground-state conductance for the Anderson Hamiltonian. Unfortunately, this approach
has only been proven successful in the region where it must give satisfactory results by
construction \cite{2013KuS030601,2017KuS}.  

To propose an alternative static approach, we have combined concepts drawn from
renormalization-group theory with the notion that the thermal dependence of the SET
conductance is parametrized by the ground-state expectation value for the quantum-dot
occupation. Chiefly important in this context is the progressive formation of the
screening cloud in the Kondo regime as the Hamiltonian crosses over from a high- to a
low-energy fixed points. At high temperatures, the dot possesse a magnetic
moment. At low temperatures, the dot spin forms a singlet with the conduction electrons.

Renormalization-group theory associates the high-energy spectrum of the Anderson Hamiltonian with
the many-body spectrum of the local-moment fixed-point Hamiltonian $H_{LM}^{*}$, and the low-energy
spectrum with that of the frozen-level fixed-point $H_{FL}^{*}$. At intermediate energies, which
correspond to the temperature range over which the dot magnetic moment is screened, the model
Hamiltonian crosses over from the vicinity of $H_{LM}^{*}$ to the vicinity of $H_{FL}^{*}$.

The physical properties describing the SET crossover are universal. In particular, as discussed
in Section~\ref{sec:20}, the electrical conductance maps linearly onto a universal function of the
temperature scaled by the Kondo temperature $T\sub{K}$. The mapping is controlled by the dot
occupancy $n\sub{d}$.

The linear mapping fits the experimental data by Grobis \etal\ \cite{GRP+08:246601} with very small
deviations. Nonetheless, the resulting dot occupancies $n\sub{d}$ are substantially lower than unity,
and the asymmetry index of the device $\kappa$ is gate-voltage dependent and exceeds unity.

The thermal dependence of the conductance for the Anderson model in the Kondo regime
reproduces the experimental data very well, while non-universal aspects of the same model offer a
blurred picture of the SET constructed by Grobis \etal\ \cite{GRP+08:246601}. \emph{Ab initio}
treatment of the device is therefore necessary before quantitative description of the experimental
data becomes possible. Given that universality simplifies the description of the crossover from the high- to the
low-temperature fixed points, \emph{ab initio} description of the high-energy fixed point will
suffice.

From a practical viewpoint, this is convenient, for in contrast with the crossover the high-energy
region yields to perturbative treatment \cite{Wi75:773}. Moreover, the high-energy fixed point
having no characteristic energy scales, its properties are temperature independent. The
ground-state energy of $H_{LM}^{*}$ can therefore be computed by standard DFT methods.

By contrast, the crossover to the $FL$ fixed point calls for special,
non-perturbative mathematical procedures. In this context, only the
the Bethe-Ansatz \cite{TW83:453} and the numerical
renormalization-group \cite{CHZ94.19,BCP08:395,YSO2009:235317}
approaches have yielded exact of essentially exact results. The
remaining challenge is to adapt one of those two methods, so that the
DFT treatment of the high-energy spectrum can serve as input for the
non-perturbative description of the crossover. Current work is
addressing that problem \cite{2018ZaOunp}.

\section{Acknowledgments}
\label{sec:30}

Three decades ago, Hardy Gross showed the ropes of DFT to one of us
(LNO). In exchange, he requested a talk on the Kondo problem, which
was never materialized. So here it is, Hardy, at long last. Too
little, and too late, but drawn from the heart, in memory of the grand
time. Close your eyes, for a second, and you may see Walter and the
two of us gathered around his table, lit by the setting sun and a
hundred flickering rays coming from the blue waters of the
Pacific. Thanks for those moments, and for all that I have learned
from you. Live long, Hardy, and leads us farther.

We are grateful to Mike Grobis for sending us the experimental data and thank Harold
Baranger and Vivaldo Campo for very helpful discussions at an early stage of this
work. The FAPESP (grant 2012/02702-0) and CNPq (grants 312658/2013-3 and 401414/2014-0)
have supported the research.  

\bibliographystyle{unsrt} 
\bibliography{epjb}

\end{document}